# PRICE INCREASES FROM ONLINE PRIVACY

by Michael R. Ward and Yu-Ching Chen

ABSTRACT

Consumers value keeping some information about them private from potential marketers. E-commerce dramatically increases the potential for marketers to accumulate otherwise private information about potential customers. Online marketers claim that this information enables them to better market their products. Policy makers are currently drafting rules to regulate the way in which these marketers can collect, store, and share this information. However, there is little evidence yet either of consumers' valuation of their privacy or of the benefits they might reap through better target marketing. We provide a framework for measuring a portion of the benefits from allowing marketers to make better use of consumer information. Target marketing is likely to reduce consumer search costs, improve consumer product selection decisions, and lower the marketing costs of goods sold. Our model allows us to estimate the value to consumers of only the latter, price reductions from more efficient marketing.

## I. Introduction

The Internet has brought thousands of businesses and millions of consumers to an electronic marketplace. Moreover, the amount of retailing conducted online is expected to grow at a phenomenal rate (Ward, 2001). One of the key factors drawing these transactions away from traditional offline markets is the dramatically increased amount of information available to guide consumers' purchase decisions. Most consumers are much better able to research products, choose desired features, and compare prices than ever before. Likewise, sellers are better able to discern individual Internet users' product preferences by tracking their web browsing behavior, online store searches, and final purchasing patterns. That retailers have been better able to surreptitiously collect and exploit detailed information about individuals has been the cause for concern about the loss of privacy on the Internet. Furthermore, these concerns will likely grow as Internet usage and online retailing grow in volume and importance and Internet information gathering techniques become more sophisticated.

Retailers value this information because it better enables them to target their advertising to those potential customers who are most likely to value their products. Target marketing can reduce retailers' customer acquisition costs, improve customers' purchase decisions, and reduce customers' transactions costs. While these outcomes generate distinct benefits for the efficient functioning of markets, there is a fear that collection of this information is an invasion of the privacy of consumers. While we acknowledge that these privacy concerns are real and important, we do not address them in this paper. Instead, we focus on the consequences of limiting retailers access to consumer information.

The desirability of protecting the privacy of online shoppers represents a tradeoff between the potential benefits of preserving consumers' privacy and the benefits of better functioning markets. Markets may function better for three reasons. First, consumers' transactions costs, in terms of time spent and effort expended, may be reduced because the receive information from advertisements that is more relevant to the actual decisions they face. Second, consumers' decision making ability may be enhanced because the information in the advertisement is both more relevant to their purchase decisions and may not have been obtained by other means. Third, retailers' costs, and perhaps prices, may be reduced if they can better target their messages to consumers who have a greater likelihood of valuing the information and acting on it to purchase their product.

This paper provides estimates of the third potential benefit, lower costs and prices due to better target marketing. First, we present a brief review of the affect of information on markets and a description of the use of consumer information online. Second, we present a model that demonstrates how better targeting of advertisements can lower retailers' customer acquisition costs and consumer prices. Third, in a series of simulations, we show how the magnitude of the price decrease depends on certain key parameters. Finally, we conclude with a discussion of what the likely values of these key parameters, and thus the expected price decrease, are.

## II. Background

Stigler (1961) was the first to show that advertising, as information, will tend to decrease the mean and dispersion of prices paid. He demonstrated that, holding production costs constant, advertising can lead to more efficient consumer decisions, and thus lower prices. Providing information about products and services to the consumer aids in economizing the consumer search and helps locate low-priced sellers more readily. In this way, advertising may lower prices. Subsequently, Steiner(1973) and Benham (1972) showed that advertising can increase retailer scale to more efficient production levels. Although the consumer pays for advertising, as reflected in the price, greater scale economies lead retailers charge lower prices. It may also lower prices when sellers or producers economize on other merchandising costs and take advantage of economies of scale.

Farris and Albion (1980) formally model advertising as equal to information. Consumers can benefit from advertising simply because ads educate the consumer by revealing more information, thereby increasing price sensitivity. This, in turn, lowers both prices and monopoly power. The model shows that the consumer search for information about a product is reduced by advertising since it serves to announce a product's existence and major attributes (Nelson 1974, 1975).

Target marketing has been around for years. These methods not only provide chances for online retailers to understand their current customers, but also help to locate a new customer base. If the retailer is successful in reducing their marketing and advertising costs, it is possible for them to offer savings and other benefits to the consumer. At one end, Nielsen and IRI track the demographics and size of various media so that advertisers can place advertisements in media best suited for their clientele. Direct mail retailers take this a step further by basing promotional decisions upon information about individuals and families rather than groups. Online advertising takes this progression another step by using much richer information on individuals' characteristics and behaviors and by doing so in "real time."

Compared to other marketing disciplines, the effectiveness of direct marketing is represented by targeting and accountability (Reed, 1999). Direct marketing, or database marketing, correlates information about consumers with their expected purchasing habits. In this way, advertising effort can be directed toward those most likely to purchase the advertised product and not "wasted" on "false leads." The techniques were primarily developed for direct mail marketing in order for retailers to cut down on expenses and increase profits (Zahavi and Levin, 1997). Improved target marketing is largely responsible for the growth in direct mail expenditures has increased of 215% since 1983, compared to 126 % for all other media (Bell and Francis, 1995). The quality of the database determines the success of any direct mail campaign. With better targeting, customers receive advertising material more relevant to their purchase decisions, which in turn improves advertiser's response rates.

Emerging electronic channels have intensified the informational issues in online markets. A drastic reduction in search costs for products and product-related information should lead to more efficient consumer decisions (Lynch and Ariely, 2000). As a result, electronic marketplaces reduce the inefficiencies caused by buyer search cost, in the process reducing the ability of sellers to extract monopolistic profits while increasing the ability of markets to optimally allocate productive resources (Bakos, 1997). Retailers can expect that lower consumer search costs will intensify competition (Smith, Bailey, and Brynjolfsson, 2000).

At the same time, the plethora of product choices available to consumers makes the processing of this information into product comparisons a daunting task for the typical consumer. While price information is more readily available, there is evidence that signals of product quality and features will remain important online (Lynch and Ariely, 2000, Ward and Lee, 2000). It is not enough to have low prices and a web site, retailers must inform customers that they exist and their product matches consumers preferences. For many products this will be true only for a small niche. Ideally, retailer will wish to inform only the niche that are actually interested in the product, that is, they wish to target their marketing effort.

Internet advertising could be described as any form of commercial content such as electronic advertisements (e.g., billboards, banner ads) or corporate web sites (Du-Coffe, 1996). The advantage of Internet advertising is it helps businesses inform consumers about products or services. " Internet advertising can be delivered via any channel (e.g., video clip, print or audio), in any form (e.g., an e-mail message or an interactive game), and provide information at any degree of depth (e.g., a corporate logo or an official web site)" (p.36).

An efficient Internet advertising and electronic marketplace have made competition in E-commerce more intensify. Marketers gather consumer information from their web side to target customers for their products.

Target marketing is playing an increasingly important role in e-commerce. Unlike direct mail, Internet advertising allows for much finer pieces of information (so-called "granular-level") to guide advertising decisions. Fletcher (1995) indicates the several unique advantages offered on the web over conventional advertising: 1)Advertisements are accessible 24 hours a day, 365 days a year, 2) Web pages are accessed because of consumer interest, 3) Response and results are immediately measurable interactively, 4) Web access is free to Internet users, 5) There are no distribution or printing fees, so millions of consumers reached costs the same as one, 6) Costs are the same whether the target audience is on the other side of the planet or on the same city block, and 7) Materials and data can be updated, supplemented or changed at any time and are therefore always up-to-date ( p.1).

Retailers can collect, store, transfer, and analyze data from consumers who visit their Web Sites. Banner ads and direct mailings are two major types of Internet advertising that are used in consumer data gathering, but consumer information can be gathered in many ways. For example, when a consumer makes a purchase online, fills out a survey, or customizes a web page (e.g.myYahoo) they identify themselves and their online behaviors to marketers. These methods usually involve the use of "cookies," small files that web sites place on consumers' computers to identify them and to track their actions. When consumers click banner ads or go to certain web pages, a "cookie" my be placed by the retailer or by third party ad placement bureau. In this way, ad placement bureaus can collate information from users' visits across many web sites. According to the FTC report (2000), most sites in a random sample allowed the placement of "cookies" by third parties.

The increase of this kind of data collection has raised public concern about online privacy. In recent years, the Federal Trade Commission has advocated protecting consumer online privacy and endorsed an industry self-regulatory agreement online to restrain online company profiling by using consumer online data collected or exchanged (FTC, 2000). According to the Internet Advertising Bureau (1999 & 2000), Internet advertising expenditure reached as much as $4.6 billion in 1999, increasing by 141% compared to $1.9 billion reported in 1998. Some forecasts put online advertising at $15 billion by 2006.

This kind of data collection has increased consumer concern about online privacy. Recent surveys show that 92% of consumers are concerned about the misuse of their private information collected online (Westin, 1999). The Federal Trade Commission, the primary US government agency concerned with protecting consumer online privacy, has adopted the core principles: Notice, Choice, Access and Security. Recently, the FTC has endorsed the self-regulatory agreement made by online advertisers. "The plan, which will take effect immediately, will require web advertising companies to notify consumers of their Internet profiling activities and give customers the chance to choose whether information about their web surfing can be gathered anonymously. The companies would also promise to give consumers 'reasonable access' to personally identifiable information collected about them and make 'reasonable efforts' to protect the data they collect" (New York Times, 2000, p.1-2).

However, privacy concerns may prevent the development of E-commerce. One study estimates that privacy concerns may have caused losses of online retail sales by as much as $2.8 billion in 1999 (Forrester, 2000). In a study sponsored by the Direct Marketing Association, Turner (2001) indicates that catalog and Internet apparel consumers will pay about $1 billion in "information tax" if data restrictions are placed on information gathering by third party companies. The result may lead retailers to pass on an increased cost ranging from 3.5 to 11 percent, which is a considerable economic loss in a $15 billion market. Even if these figures were inflated, data restrictions could seriously jeopardize the viability E-commerce.

Most recent studies discuss either consumer attitudes toward online privacy, or current policy issues in the practice of online privacy. For example, Sheehan (1999) investigates whether gender differences are apparent in the attitudes and behaviors toward online privacy. However, Turner's (2001) study is the only one we know that discuss the impact of data restrictions on consumer costs. It fills a gap by addressing questions of whether online direct marketing is more efficient than other marketing strategies and whether these efficiencies will be pass on to consumers in the form of lower prices.

## III. Model

The general methodological strategy is to model firms' behaviors without being able to engage in target marketing and with being able to engage in target marketing. By target marketing we mean adjusting advertising intensity according to the expected purchasing patterns of different consumer groups. Target marketing of this sort requires detailed information that could be used to predict purchasing patterns. Stronger privacy regulations limit this sort of information and thus foreclose target marketing as an option, or weaken its effectiveness.

Our model is essentially one of monopolistic competition. An individual firm has some market power, that is, it face a downward sloping demand curve, possibly due to product differentiation. In addition, the model incorporates imperfect consumer information. Consumers do not know of a products existence or desirability until they receive an advertising message from the producer. Consumers are heterogeneous both in the utility they derive from the product and in the likelihood they will observe a message. Therefore, the two conditions necessary for a consumer to purchase are that doing so maximizes his utility and that he knows that it does.

*Consumers*

Each consumer can purchase one unit or zero. Let $U^i(x, I^i - P \times x)$ be consumer $i$'s utility reduced form function where the product in question is $x$. Purchasing ($x=1$) at price $P$ reduces his consumption of all other goods to his income less expenditures on $x$. He will consume if, and only if, $U^i(x, I^i - P \times x) > U^i(0, I^i)$. Consumers are heterogeneous in their marginal utility of $x$ and all other goods so that not all consumers would purchase good $x$ at $P > 0$. At lower prices, however, more consumers will purchase x because their purchases of all other goods are not reduced as much. For our purposes below we need only define $\pi(P) = prob(U^i(x, I^i - Px) > U^i(0, I^i))$, or the probability of purchasing $x$. In a more general model, $\pi(P)$ would also be a function of competing products' prices. For reasons noted below, we do not wish to, nor need to, model industry structure.

Suppose that it is possible to identify two groups such that $\pi_1(P) > \pi_2(P)$, the probability that any consumer in group one will purchase good *x* is non-trivially higher that any consumer in group two, all else equal. We assume that target marketing identifies individuals as being in one group or another while, without target marketing, retails cannot tell what group an individual belongs to. In this case, they only know $G(P) = w_1 \pi_1(P) + w_2 \pi_2(P)$ where $w_i$ is the fraction of people in each group.

*Firms*

Firms bear production costs and advertising costs. Production costs are summed to be constant marginal costs *C* and fixed costs *F*. Advertising costs increase linearly with advertising intensity, *A*, and the population, *N*. Demand increases with advertising, *A*, and decreases with price, *P*. Therefore, profits are given by $(P-C)Q(P,A) + F + RAN$. Since firm level demand is higher at lower prices, firms have market power and earn rents, which turn out to be quasi-rents. These quasi-rents just cover their fixed costs and advertising costs. Entry and exit force firms to earn zero expected economic profits in the long run.

*Information*

In order to purchase, consumers must know both that the product exists and has features such that his utility is higher with consumption. Consumers usually are not fully informed, however. The fraction who are informed increases with advertising intensity, *A*. Define $\phi(A)$ as the fraction of consumers who are informed. This is assumed to increase with advertising per person at a decreasing rate, $\phi' > 0$ and $\phi'' < 0$ (see Figure 1). Then the probability of purchase is $\phi(A)\pi(P)$ and the quantity of the firm's product demanded by consumers, $Q(P,A)$, is given by $N((A)''(P)$. Therefore, profits are given by:

$$\mathbf{P} = (P-C)N(A)a(P) - F - RAN.$$

*Equilibrium*

Equilibrium is characterized by three conditions. First, marginal profits with respect to price are zero. This yields the familiar first order condition, $MA/MP = 0$, for interior solutions. Second, marginal profits with respect to advertising are zero. This yields the first order condition $MA/MA = 0$ for interior solutions. Third, economic profits are zero in long-run equilibrium, $A = 0$. Typically, these three conditions allow us to identify three endogenous parameters: price, advertising and industry structure. That is, typically, demand is a function of an industry structure parameter, like the number of firms. Positive profits cause entry that shifts in each firm's demand curve and reduces profits.

We are only interested in price and advertising and not in industry structure. Moreover, modeling this usually requires strong assumptions on how firm level demand is affected by entry. For example, with more firms, does an individual firm's residual demand curve shift in horizontally or does it rotate? Similarly, imposing the first condition, $MA/MP = 0$, usually requires strong assumptions on the shape of demand curves. Instead, we solve for just $P$ and $A$ from the second and third conditions and assume that industry structure adjusts so as to restore the fist condition.

*Profit Maximization*

The firm has two choice variables, $P$ and $A$. We assume there is an interior solution so that first order conditions apply. While we do not use this condition below to solve for price effects, it is still informative to see what it implies. Differentiating profits with respect to price yields:

$$Ng(A)a(P) + (P - C)Ng(A)a'(P) = 0.$$

Using this condition requires that we know the shape of consumer demand. Specifically, we must approximate $a(P)$ and $a'(P)$. These functions are likely to be quite different depending on the application. Below, we choose not to specify $a(P)$ and so do not use this condition. Still, we can use this equation to generate the familiar Lerner index. Recall that $Ng(A)a(P)$ is the specification of $Q(P,A)$. Then, $Ng(A)a'(P)$ is given by $Q_P(P,A)$. Dividing by, $Q(P,A)$ and multiplying the second term by $P/P$ yields:

$$1 + \frac{h(P - C)}{P} = 0 \text{ or } \frac{P - C}{P} = \frac{-1}{h}$$

where $h$ is the own-price elasticity.

Differentiating profits with respect to advertising level yields:

$$(P - C)g'(A)a(P) = R$$

or the value of the marginal advertising expense equals its cost. Rearranging yields advertising intensity as a function of other parameters:

$$A = g'^{(-1)}\left(R / [a(P)(P - C)]\right).$$

The zero profit condition implies that:

$$(P - C)Ng(A)a(P) = F + RAN$$

or the rents earned on prices above marginal costs are just offset by fixed production and advertising costs. Below, we will approximate $a(P)$ with the constant $a$ for small changes in $P$. In this case, the above expression can be solved for price as:

$$P = C + (F + RAN) / Ng(A)a(P)$$

or price equal average cost. Together, equations () and () define a system of two equations and two unknowns. The most difficult aspect of solving these equations is the form of $\epsilon(A)$.

*Gains to Target Marketing*

For the most part, equations () and () define the equilibrium without target marketing. Simply replace $\eta$ with $G$ in equation () to find $A^*$. Next, solve () for $P^{w/oTM}$ using this value of $A^*$. This method assumes that $\eta$ is not a function of price in contradiction to the assumptions of the general model. This will yield valid approximations for small changes in $P$, and thus $\eta(P)$.

A firm engaging in target marketing now can choose different advertising intensity levels for the two groups. She maximizes:

$$\Pi = (P - C)\left[a_1 N_1 g(A_1) + a_2 N_2 g(A_2)\right] - R_1 A_1 N_1 - R_2 A_2 N_2 - F$$

with respect to $P$, $A_1$ and $A_2$. The first order conditions with respect to advertising are:

$$g'(A_1) = R_1 / a_1(P - C) \text{ and } g'(A_2) = R_2 / a_2(P - C).$$

In order to use target marketing, the relative price for the advertising price for the target audience must not be as high as it relative productivity, or $R_1/R_2 < \eta_1/\eta_2$. In this case, $g'(A_1) < g'(A_2)$. Since there is diminishing marginal returns to advertising, this requires $A_1 > A_2$.

Since $\eta_1 > G > \eta_2$, we have that $g'(A_1) < g'(A^*) < g'(A_2)$ and therefore $A_1 > A^* > A_2$. This implies that $A_1(A_1) > A_1(A^*)$ and $A_2(A_2) > A_2(A^*)$ and therefore $A_1(A_1) + A_2(A_2) > A(A^*)$. That is, target marketing is increase profits.

These are short-run profits. In the long run, profits will entice entry causing price to fall from the initial $P^{w/oTM}$ to $P^{TM}$ so as to restore the zero profit condition, equation (). The difference between $P^{w/oTM}$ and $P^{TM}$ represents a consumer gain from more effective target marketing. For various parameter values and assumptions, we can calculate *($P^{w/oTM}$ - $P^{TM}$)/$P^{w/oTM}$* as percentage price reduction.

## IV. Numerical Examples

In order to simulate the effects of target marketing on costs, prices and welfare, we first need to assume a specific functional form for $\Gamma$, the information production function. For the most part, any $\Gamma$ that satisfies certain conditions will suffice. These conditions are: a) $\Gamma(0) = 0$, b) $\Gamma'(A) > 0$, c) $\Gamma''(A) < 0$, and d) $\Gamma(\infty) = 1$. Together, these imply that $\Gamma'(\infty) = 0$. In addition, if we assume $\Gamma'(0) = \infty$, we are guaranteed an interior solution to (). In particular, we derive such one such $\Gamma$ from first principles and assumptions on information production.

Assume that the probability that a consumer sees any message is the constant $\pi$. If messages are independently and identically distributed, then the number of messages a consumer sees follows a binomial distribution. A consumer is considered informed if he sees at least one message. Under these assumptions, the probability of not seeing message after $M$ trials is $(1-\pi)^M$. Suppose further that advertising intensity increases messages at a decreasing rate such that, $M(A) = \sqrt{A}$. Under these assumptions, $\Gamma(A) = 1-(1-\pi)^{\sqrt{A}}$. This function satisfies all the conditions above. For future reference, note that $\Gamma'(A) = (1-\pi)^{\sqrt{A}} \ln(1-\pi)[1/(2\sqrt{A})]$.

We can now apply () and () to determine the effect of using target marketing for specific parameter values. Assume, $\sigma_1 = 0.4$, $\sigma_2 = 0.04$, and $N_1 = N_2 = 500$, making $\Theta = 0.22$. Assume further that $C = \$8$, $R_1 = \$0.0100$, $R_2 = \$0.0125$, and $F = \$50$. Finally, assume, $\pi = 0.10$. Under these assumptions, $A^* = 4.06$ and $Q^*=42.08$. Since there are 220 potential customers, $\theta N$, this value of $Q^*$ implies a consumer take up rate of 19.1%. At $A = 0$, $P^{w/oTM} = \$10.152$. Similarly, $A_1 = 6.13$, $A_2 = 0.09$, $Q_1 = 45.92$, and $Q_2 = 0.62$. Note that a ten-fold increase in the probability of purchase, $\theta$, yielded a 68-fold increase in advertising intensity and a 74-fold increase in the probability of purchase. Since there are 200 potential customers in group 1, $\theta_1 N_1$, and 20 in group 2, $\theta_2 N_2$, these represent consumer take up rates of 23.0% and 3.1%. At $P = \$10.152$, the price charged without target marketing, the firm would earn \$11 in profit on about \$513 revenue. With free entry, price would fall to $P^{TM} = \$9.907$ a more than 2.4% decrease in price due to target marketing.

These particular parameter values were picked somewhat arbitrarily, but yield ratios not far from commonly reported. Without target marketing, advertising expense represented 9.5% of sales and fixed costs represented 11.7% of total costs. With target marketing, advertising expense represented 8.4% of sales and fixed costs represented 10.8% of total costs. While actual estimates of fixed costs in the long run are hard to find advertising-to-sales ratios of 8-10% are found in moderately heavily advertising industries (Nelson, 1974).

Moreover, from these values we can calculate *P-M/P* which, in equilibrium, is equal to the inverse of the own-price elasticity, $\eta$. In this case, we calculate elasticities of -4.7 and -5.2 without and with target marketing. These are within the range of those reported for many branded consumer products (Hausman and Leonard, 1997).

*Simulations*

Table 1 reports the equilibrium values as the size of group 1 falls. Group 1 corresponds to the group with the higher likelihood of purchasing. Above, the two groups were assumed to be of equal size, implying that $w_1$ was 50%. In Table 1, $w_1$ is successively reduced to 25%, 10% and 5%. Consequently, the fraction of the population in Group 2, $w_2$, is increased to 75%, 90% and 95%. In general, the gains to target marketing steadily increase as $w_1$ falls reaching a 12% price reduction when $w_1$ reaches 0.05%.

Table 2 reports the equilibrium values as the difference in the purchase probabilities between the groups falls. In the base case $\pi_1$ is assumed to be 0.4, while $\pi_2$ is assumed to be just 0.04. This represents a tenfold increase in the purchase probability based group membership alone. It is not clear that target marketing will reveal such large differences in purchasing behavior (but see ...). To gauge how sensitive the price change results are to this assumption, Table 2 reports equilibrium values as the difference between $\pi_1$ and $\pi_2$ decreases, while $G$ remains constant. The price decrease due to target marketing steadily falls as the groups become more similar. In fact, when $\pi_1$ equals 0.28 and $\pi_2$ equals 0.16, prices are actually higher with target marketing. Intuitively, this is because the difference in advertising productivity, from $G = 0.22$ to $\pi_1 = 0.28$, is no longer worth the higher price for targeted advertising. One could argue that increased privacy limits the just sort of information that allows marketers to better classify consumers by probability of purchase.

Table 3 reports the equilibrium values for larger fixed product costs, $F$. Essentially, prices fall because, with more efficient advertising, these fixed costs can be spread over more sales (see equation ()). Table 3 investigates how sensitive results are to fixed costs by doubling $F$ from $50 to $100. Relative to the comparable columns in Table 1, Table 3 yields slightly larger price decreases. Note, however, that in the last column price is more than double marginal costs, or, by the Lerner index, the own-price elasticity, $O$, is between -2 and -1. This is likely to be more inelastic than exists for most consumer products.

Table 4 reports the equilibrium values when advertising is more productive, $\lambda$ doubles to 0.20. This results in consumers having a higher probability of observing each advertising message. This tends to make $\zeta$ steeper initially and flatten out sooner. Because of this, advertising reaches diminishing returns sooner, resulting in a general decrease in the amount of advertising. Since advertising levels are reduced, there are fewer fixed costs to spread across units and price levels are uniformly lower. Moreover, advertising efficiency increases with target marketing with smaller reductions in advertising expense relative to Table 1. Therefore, price decreases are about half as large as in Table 1.

In all these simulations, the calculated price changes stem from the difference between the use of target marketing or not. However, this is not the decision facing policy makers. The correct simulation would compare "some" target marketing with "more" target marketing. For example, what is the gain from identifying ten groups of consumers over identifying eight? Or, what is the gain from identifying groups with more disparate purchase patterns ($\pi_1 = .4$ and $\pi_2 = .04$ versus $\pi_1 = .28$ and $\pi_2 = .16$)? We are reluctant to undertake such simulations until we know better what the key parameters are, specifically $\zeta(A)$ and the change in $\pi_i$'s due to target marketing. We admit that the parameter values that we have chosen are, at best, educated guesses. Nevertheless, they yield non-trivial price changes. To better refine these estimates, we need empirical estimates of these likely changes.

## V.  Conclusions

Many of our simulations can yield plausible price changes greater than 1%, some much greater. This represents a real benefit to consumers in the form of greater consumer surplus. For example, since online retailing represented about $40 billion in the US in 2000, a 1% higher cost due to better privacy

would have come to $0.4 billion. Since online information can lead to as many offline purchases as online (Ward and Morganosky, 2001), the cost could have reached $0.8 billlion. Moreover, since online retailing is expected to increase five-fold over the next four years (Ward, 2001), the cost would come to $4 billion by then. Even for just a 1% price change, these are not inconsequential amounts.

This paper demonstrates that protecting privacy represents a difficult tradeoff. We explore one mechanism for consumer costs from protecting their privacy. We attempt to gauge the magnitude of these costs. The best we can do at this point is indicate very general magnitudes. However, with better estimates of the effectiveness of online target marketing, the model we propose could yield precise estimates of the price increases due to online privacy.

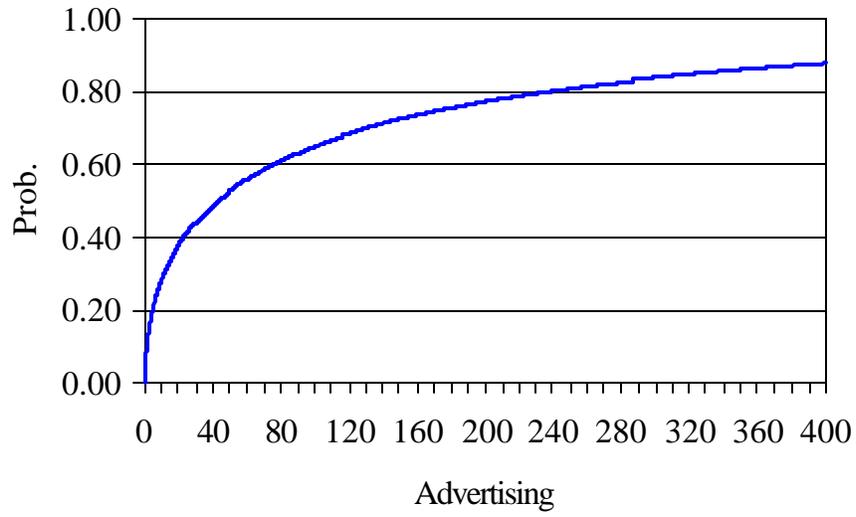

Figure 1
Probability of Being Informed (Gamma)

Table 1
Simulated Effects of Group Size
Base Case with various values of $w_1$

| | | | | |
|---|---|---|---|---|
| Fraction in Group1 - $w_1$ | 0.500 | 0.250 | 0.100 | 0.050 |
| Group 1 Probability of Purchase - $\pi_1$ | 0.400 | 0.400 | 0.400 | 0.400 |
| Group 2 Probability of Purchase - $\pi_2$ | 0.040 | 0.040 | 0.040 | 0.040 |
| Average Probability of Purchase - $\bar{\pi}$ | 0.220 | 0.130 | 0.076 | 0.058 |
| Advertising w/o Target Marketing - $A^*$ | 4.060 | 4.060 | 4.060 | 4.060 |
| Units Sold w/o Target Marketing - $Q^*$ | 42.080 | 24.866 | 14.537 | 11.094 |
| Price w/o Target Marketing - $P^{w/oTM}$ | $10.152 | $11.643 | $14.232 | $16.166 |
| Advertising for Group 1 - $A_1$ | 6.130 | 10.920 | 21.580 | 32.720 |
| Advertising for Group 2 - $A_2$ | 0.140 | 0.300 | 0.740 | 1.330 |
| Units Sold for Group 1 - $Q_1$ | 45.922 | 29.402 | 15.481 | 9.053 |
| Units Sold for Group 2 - $Q_2$ | 0.773 | 1.682 | 3.119 | 4.348 |
| Price with Target Marketing - $P^{TM}$ | $9.907 | $10.778 | $12.496 | $14.200 |
| Percentage Price Change | -2.4% | -7.4% | -12.2% | -12.2% |

Assumes marginal costs, $C$, are $8, fixed costs, $F$, are $50, advertising costs, $R$, are $0.01 per person without target marketing and $0.0125 per person with target marketing, population, $N$, is 1,000, and the probability of observing a message, $\gamma$, is 0.10.

Table 2
Simulated Effects of Group Differences
Base Case with Various Values of $\pi_1$ and $\pi_2$

| | | | | |
|---|---|---|---|---|
| Fraction in Group1 - $w_1$ | 0.500 | 0.500 | 0.500 | 0.500 |
| Group 1 Probability of Purchase - $\pi_1$ | 0.400 | 0.380 | 0.340 | 0.280 |
| Group 2 Probability of Purchase - $\pi_2$ | 0.040 | 0.060 | 0.100 | 0.160 |
| Average Probability of Purchase - $\bar{\pi}$ | 0.220 | 0.220 | 0.220 | 0.220 |
| Advertising w/o Target Marketing - $A^*$ | 4.060 | 4.060 | 4.060 | 4.060 |
| Units Sold w/o Target Marketing - $Q^*$ | 42.080 | 42.080 | 42.080 | 42.080 |
| Price w/o Target Marketing - $P^{w/oTM}$ | $10.152 | $10.152 | $10.152 | $10.152 |
| Advertising for Group 1 - $A_1$ | 6.130 | 6.000 | 5.580 | 4.490 |
| Advertising for Group 2 - $A_2$ | 0.140 | 0.340 | 1.000 | 2.560 |
| Units Sold for Group 1 - $Q_1$ | 45.922 | 43.219 | 37.456 | 28.012 |
| Units Sold for Group 2 - $Q_2$ | 0.773 | 1.788 | 5.000 | 12.411 |
| Price with Target Marketing - $P^{TM}$ | $9.907 | $9.981 | $10.116 | $10.247 |
| Percentage Price Change | -2.4% | -1.7% | -0.4% | 0.9% |

Assumes marginal costs, $C$, are $8, fixed costs, $F$, are $50, advertising costs, $R$, are $0.01 per person without target marketing and $0.0125 per person with target marketing, population, $N$, is 1,000, and the probability of observing a message, $\gamma$, is 0.10.

Table 3
Simulated Effects of Fixed Costs
Base case with $F = \$100$

| | | | | |
|---|---|---|---|---|
| Fraction in Group1 - $w_1$ | 0.500 | 0.250 | 0.100 | 0.050 |
| Group 1 Probability of Purchase - $\pi_1$ | 0.400 | 0.400 | 0.400 | 0.400 |
| Group 2 Probability of Purchase - $\pi_2$ | 0.040 | 0.040 | 0.040 | 0.040 |
| Average Probability of Purchase - $\bar{\pi}$ | 0.220 | 0.130 | 0.076 | 0.058 |
| Advertising w/o Target Marketing - $A^*$ | 7.570 | 7.570 | 7.570 | 7.570 |
| Units Sold w/o Target Marketing - $Q^*$ | 55.363 | 32.715 | 19.125 | 14.596 |
| Price w/o Target Marketing - $P^{w/oTM}$ | $11.173 | $13.370 | $17.186 | $20.037 |
| Advertising for Group 1 - $A_1$ | 11.240 | 19.560 | 37.060 | 40.000 |
| Advertising for Group 2 - $A_2$ | 0.310 | 0.650 | 1.600 | 2.840 |
| Units Sold for Group 1 - $Q_1$ | 59.517 | 37.248 | 18.938 | 9.728 |
| Units Sold for Group 2 - $Q_2$ | 1.139 | 2.443 | 4.492 | 6.182 |
| Price with Target Marketing - $P^{TM}$ | $10.832 | $12.182 | $14.859 | $17.552 |
| Percentage Price Change | -3.1% | -8.9% | -13.5% | -12.4% |

Assumes marginal costs, $C$, are $8, fixed costs, $F$, are $100, advertising costs, $R$, are $0.01 per person without target marketing and $0.0125 per person with target marketing, population, $N$, is 1,000, and the probability of observing a message, $\gamma$, is 0.10.

Table 4
Simulated Effects of Advertising Efficiency
Base case with $\gamma = 0.20$.

| | | | | |
|---|---|---|---|---|
| Fraction in Group1 - $w_1$ | 0.500 | 0.250 | 0.100 | 0.050 |
| Group 1 Probability of Purchase - $\pi_1$ | 0.400 | 0.400 | 0.400 | 0.400 |
| Group 2 Probability of Purchase - $\pi_2$ | 0.040 | 0.040 | 0.040 | 0.040 |
| Average Probability of Purchase - $\bar{\pi}$ | 0.220 | 0.130 | 0.076 | 0.058 |
| Advertising w/o Target Marketing - $A^*$ | 3.380 | 3.390 | 3.390 | 3.390 |
| Units Sold w/o Target Marketing - $Q^*$ | 74.033 | 43.799 | 25.605 | 19.541 |
| Price w/o Target Marketing - $P^{w/oTM}$ | $9.131 | $9.915 | $11.276 | $12.293 |
| Advertising for Group 1 - $A_1$ | 4.930 | 8.300 | 14.910 | 20.790 |
| Advertising for Group 2 - $A_2$ | 0.170 | 0.350 | 0.860 | 1.450 |
| Units Sold for Group 1 - $Q_1$ | 78.142 | 47.422 | 23.101 | 12.770 |
| Units Sold for Group 2 - $Q_2$ | 1.758 | 3.710 | 6.729 | 8.954 |
| Price with Target Marketing - $P^{TM}$ | $9.021 | $9.536 | $10.560 | $11.533 |
| Percentage Price Change | -1.2% | -3.8% | -6.3% | -6.2% |

Assumes marginal costs, $C$, are $8, fixed costs, $F$, are $50, advertising costs, $R$, are $0.01 per person without target marketing and $0.0125 per person with target marketing, population, $N$, is 1,000, and the probability of observing a message, $\gamma$, is 0.20.